\documentclass[draft]{agujournal}
\draftfalse

\usepackage{hyperref}
\usepackage{rotating}

\journalname{Space Weather}

\begin{document}

%
%


\title{Forecasting the Arrival Time of Coronal Mass Ejections: Analysis of the CCMC CME Scoreboard}

%
%




\authors{
Pete Riley\affil{1}, Leila Mays\affil{2}, Jesse Andries\affil{3}, Tanja Amerstorfer\affil{4},  Douglas Biesecker\affil{5}, V\'eronique Delouille\affil{3}, Mateja Dumbovi\'{c}\affil{6,7}, Xueshang Feng\affil{8}, Edmund Henley\affil{9}, Jon A. Linker\affil{1}, Christian M\"{o}stl\affil{4}, Marlon Nu\~{n}ez\affil{10},Vic Pizzo\affil{5}, Manuela Temmer\affil{4}, W.K. Tobiska\affil{11}, C. Verbeke\affil{12}, Matthew J West\affil{3}, and Xinhua Zhao\affil{6}
}
\affiliation{1}{Predictive Science Inc., San Diego, USA}
\affiliation{2}{NASA/GSFC, Greenbelt, MD 20771, USA}
\affiliation{3}{Solar-Terrestrial Center of Excellence, Royal Observatory of Belgium, Ringlaan 3, B-1180 Brussels, Belgium}
\affiliation{4}{Space Research Institute, Austrian Academy of Sciences, 8042 Graz, Austria}
\affiliation{5}{Space Weather Prediction Center, NOAA, Boulder, Colorado, USA}
\affiliation{6}{Institute of Physics, University of Graz, Graz, Austria}
\affiliation{7}{Hvar Observatory, Faculty of Geodesy, University of Zagreb, Zagreb, Croatia}
\affiliation{8}{SIGMA Weather Group, State Key Laboratory of Space Weather, National Space Science Center, Chinese Academy of Sciences, Beijing 100190, China}
\affiliation{9}{Met Office, FitzRoy Road, Exeter, Devon, UK}
\affiliation{10}{Department of Languages and Computer Sciences, Universidad de M\'{a}laga, Málaga, Spain}
\affiliation{11}{Space Environment Technologies, Pacific Palisades, CA 90272, USA}
\affiliation{12}{Centre for Mathematical Plasma-Astrophysics, KU Leuven, Leuven, Belgium}





\correspondingauthor{Pete Riley}{pete@predsci.com}




\begin{keypoints}
\item Overall, current forecasts of the arrival time of CME-driven shocks have mean errors of $\pm 10$ hours, with standard deviations of $\pm 20$ hours.  
\item The most accurate model can forecast the arrival time of CME-driven shocks with a mean error (bias) of -1 hour,  a mean absolute error of 13 hours, and standard deviation of 15 hours.  
\item Arrival time forecasts have not improved in accuracy during the previous six years.  
\end{keypoints}

%
%


\begin{abstract}
Accurate forecasting of the properties  of coronal mass ejections as they approach Earth is now recognized as an important strategic objective for both NOAA and NASA. The time of arrival of such events is a key parameter, one that had been anticipated to be relatively straightforward to constrain. In this study, we analyze forecasts submitted to the Community Coordinated Modeling Center (CCMC) at NASA's Goddard Space Flight Center over the last six years to answer the following questions: (1) How well do these models forecast the arrival time of CME-driven shocks? (2) What are the uncertainties associated with these forecasts? (3) Which model(s) perform best? (4) Have the models become more accurate during the past six years? We analyze all forecasts made by 32 models from 2013 through mid 2018, and additionally focus on 28 events all of which were forecasted by six models. We find that the models are generally able to predict CME-shock arrival times -- in an average sense -- to within $\pm 10$ hours, but with standard deviations often exceeding 20 hours. The best performers, on the other hand, maintained a mean error (bias) of -1 hour, a mean absolute error of 13 hours, and a precision (s.d.) of 15 hours. Finally, there is no evidence that the forecasts have become more accurate during this interval. We discuss the intrinsic simplifications of the various models analyzed, the limitations of this investigation, and suggest possible paths to improve these forecasts in the future. 
\end{abstract}

%
%

%


%
%
%
%

\section{Introduction}

Space weather refers to the conditions surrounding and within the Earth's environment, driven by changes in solar activity. Although it can broadly encompass a wide range of effects, of primary concern is the interaction of Coronal Mass Ejections (CMEs) with the Earth's magnetosphere, ionosphere, atmosphere, and lithosphere. The strength of this interaction is controlled to a large degree by the speed of the arriving CME, and the amount of southward pointing magnetic field ($B_z$) contained within it. 

Prediction of $B_z$ has remained a crucial, but exceedingly difficult task, in spite of great effort being expended on it \citep[e.g.][]{riley17b}. However, given that an event has been observed by remote solar observations, a related, and arguably equally important question to the ``what will hit Earth?'' is ``when will it hit Earth?'' In principle, this is a much simpler problem to solve: The complex details of the eruption process, and the evolution of the flux rope within the ejecta as it propagates through an inhomogeneous medium do not need to be solved. Instead, given a wide range of initial signatures signaling the launch of the ejecta, one needs only to identify a reasonable speed profile from $30 R_S$ to 1 AU -- for the part of the CME that propagates along the Sun-Earth line -- to estimate the time of transit of the ejecta and/or its associated shock wave. 

Many models have been developed over the years to estimate the time of travel, or arrival time, of the CME. In some cases, the shock itself is the focus of the prediction, while in other cases, it is the ejecta itself. \citet{zhao14a} provided a detailed and thorough discussion of the types of models that have been developed for, or adapted for the purpose of forecasting the time of arrival of CMEs and/or their shocks. They categorize the models as follows: empirical models, expansion speed models, drag-based models, physics-based models, and MHD models. These distinctions are relevant to the forecast submissions that we analyze here, in that each group is represented by at least a handful of models. 

In 2013, NASA's Community Coordinated Modeling Center (CCMC) developed a web-based submission form for community researchers and operational forecasters to submit their forecast for CME-driven shock arrival times, and, optionally, other space weather parameters if appropriate. The underlying philosophy was that once a new CME had been identified at the Sun, the users would submit their forecast in real-time (or as close as possible, since the lead time of the prediction was also tracked). Additionally, the forecasts would be made available to the community, again in real-time, allowing users and the community-at-large to view the forecasts as they came in. While submitting teams must be registered, the pages were open to anyone to view. Until now, no rigorous analysis of these forecasts has been made (although a preliminary analysis of a subset of these results for two specific models was performed by \citet{pope16a}). 

The CME scoreboard is a component of a broader CME Arrival Time and Impact Working Team started in 2017 and facilitated by the CCMC (\url{https://ccmc.gsfc.nasa.gov/assessment/topics/helio-cme-arrival.php}).  While the scoreboard focuses primarily on predicting CME-driven shock arrival before it is observed, the working team, in conjunction with the scientific community, will evaluate how well different models/techniques can predict arrival times and geomagnetic impacts for a set of $\sim 100$ historical events. Its goals are: (1) to evaluate the current status of CME arrival time and impact prediction; (2) to establish metrics agreed upon by the community; and (3) to provide a benchmark against which future models and model improvements can be assessed.

Various quantities can be calculated to estimate the accuracy of CME-shock arrival times. Most simply, the forecast error for a particular prediction, $i$, can be defined as: 

\begin{equation}
\Delta t_i = t_i^o - t_i^f
\end{equation}
where $t_i^o$ is observed arrival time of the $i$th CME shock and $t_i^f$ is the forecasted arrival time. Although counter to the more typical definition of anomalies, where the true value is subtracted from the forecasted value, this definition has the intuitive property that $\Delta t < 0$ implies that the CME-driven shock arrived earlier than the forecast, while  $\Delta t > 0$ implies that it arrived later.

We define the accuracy (or bias) for a number of forecasts ($N$) as the mean error: 

\begin{equation}
accuracy  = <\Delta t>  = \frac{1}{N} \sum_{i=1}^N \Delta t_i 
\end{equation}

Precision is defined as the standard deviation (s.d.), which in turn is the square root of the variance: 

\begin{equation}
s.d  = \sqrt{variance}= \sqrt{\frac{1}{N} \sum_{i=1}^N  |\Delta t_i - <\Delta t>|^2} 
\end{equation}

Finally, the mean absolute error (MAE) is defined by: 

\begin{equation}
MAE = \frac{1}{N} \sum_{i=1}^N |\Delta t_i -<\Delta t>|
\end{equation}

Although accuracy (mean error) is often reported, since positive and negative errors tend to cancel one another out, a more meaningful metric for accuracy, we believe, is the MAE \citep[e.g.][]{morley18a}. However, the mean error remains important since it conveys information about possible forecast bias, i.e., any tendency for the model to systematically under- or overestimate the observed arrival time. Thus, in this study, we report all three quantities: mean error, MAE, and s.d. Additionally, for completeness, we also calculate: minimum, first quartile, median, third quartile, and maximum values. 

The purpose of this study is four-fold. First, to explore the accuracy and precision of the predictions made by 32 teams during the previous six years. Second, to identify any trends, such as improvements in accuracy during this time period. And third, to identify any specific model, or group of models, that appear to perform better than the rest. Fourth, to provide a basis for evaluating future, novel forecast submissions by publishing these results, together with the code necessary to update the study in the future. 

\section{Methods}

\subsection{Models}

Currently, 32 distinct models (or model variants) have been used to predict the CME arrival times and shock arrival times during the interval from 2013 through late 2017. These are summarized in Table~\ref{table-models}. The models are summarized elsewhere (\url{https://swrc.gsfc.nasa. gov/main/cmemodels/}), which also provides a list of peer-reviewed papers describing the techniques. Here we illustrate a few of the approaches, focusing on the models that have made the most number of predictions, as well as those that can be conveniently compartmentalized into a particular category.  

In the broadest terms, the models can be classified as ``CME shock'' arrival forecasts or ``CME'' arrival forecasts. In the former category, the STOA (shock time of arrival) \citep{dryer04a} and WSA-ENLIL+Cone Model \citep{odstrcil04a} are two prominent examples. In the latter category, the WSA-ENLIL+Cone Model also plays a major role, while the Drag Based Model (DBM) \citep{vrvsnak13a} serves to illustrate a complementary approach to the problem. 

The STOA model is undoubtedly the longest running shock forecast model, and, arguably the simplest to implement. It assumes that a shock, generated from the eruption of a CME (which is observed in white light) has a speed profile that is initially constant, after which it decays as a blast wave, with $V_s \sim R^{-1/2}$. To address the fact that the shock is propagating through an inhomogeneous medium, the speed of the solar wind at 1 AU at the time of the flare is used to scale the evolution of the shock wave. This is a crude attempt to incorporate the ambient solar wind conditions, but does not account for stream interaction regions, and other structures that the shock may encounter on its journey from the Sun to 1 AU. The main inputs to the model are: (1) the flare's solar longitude; (2) the start time of the metric radio type II radio drift, the duration of the GOES X-ray trace (which acts as a proxy for the duration of the piston-driving portion of the velocity profile); and the solar wind speed at 1 AU at the time of the flare. The model outputs the shock arrival time, amongst other derivative parameters. 

The WSA-ENLIL+Cone Model, including its many implementations (e.g., NOAA, GSFC/CCMC, and UK Met Office) is the current standard model for predicting the large-scale plasma properties of the ejecta as it propagates from $\sim 20 R_S$ to 1 AU. It is also representative of some of the other advanced MHD-based models that are currently being, or proposed to be used for CME forecasts (e.g., CORHEL, SWMF, EUHFORIA, SUSANOO, and ZEUS-3D), and thus serves to illustrate several general points. Moreover, it is the only operational space weather model implemented by NOAA in the USA. The forecasts analyzed here were originally produced by NOAA forecasters working at the Space Weather Prediction Center (SWPC) in Boulder, CO. The model is initialized by parameters derived from a ``cone model'' fit to white-light images as the CME is observed to pass through the solar corona. Specifically, the initial speed, density, location, and propagation direction of the plasma ejecta are all derived from these observations and serve as boundary conditions for the heliospheric model, ENLIL. ENLIL, itself is first populated with ambient solar wind flow using the empirically-based WSA model. It is worth noting that WSA-ENLIL+cone Model purposefully makes some simplifications to the process that provide tractability for forecasting purposes \citep{pizzo11a}. More sophisticated models of the ambient solar wind now include thermodynamics, and even wave/turbulent heating. Additionally, more advanced CME models include the eruption process, which provides a self-consistent flux rope embedded in the CME. ENLIL, on the other hand, provides only ambient spiral fields within its ejecta. Nevertheless, for the purposes of forecasting the arrival of the CME and its shock, these are defensible simplifications.

Given the several variants of the WSA-ENLIL+Cone (WEC) models in the scoreboard, it is worth commenting on some of the  distinctions. The UK's Met Office WEC model entry, produced by the Met Office Space Weather Operations Centre (MOSWOC), for example, is based on a``human-in-the-loop'' interpretation of ENLIL output, not the raw (or automated) output. The forecaster  uses the WSA-ENLIL result as one factor in their CME arrival time prediction, manually adjusting this prediction based on their experience and any other observations. For example, if the comparison between ENLIL and {\it in situ} measurements at 1 AU suggests that the ENLIL background solar wind speed is lower than in reality, forecasters may nudge the forecast to have the CME arriving at Earth earlier than the raw ENLIL output would have suggested. Additionally, it is worth noting that the Met Office forecasts non-Earth directed events, incorporating the probability column in the scoreboard to reflect their confidence that it might result in a glancing collision with Earth.

The Drag-Based Model (DBM) lies between the STOAA and WSA-ENLIL+cone models in terms of complexity. It relies on the assumption that the dynamics of ICMEs can be interpreted by the MHD drag: ICMEs that are faster than the ambient flow are decelerated, whereas those traveling slower than the ambient wind are accelerated. \citet{vrvsnak13a} derived a set of expressions that allow the equation of motion to be solved analytically, producing, in part, the time of arrival of the CME. Assumptions must be made for several free parameters in the model, which are, in turn, based on reasonable, but average properties of CMEs near the Sun. The model relies on several important assumptions, including one that requires the mass of the ICME to remain constant. Additionally, it does not take into account the structure of the ambient solar wind into which it propagates. The primary inputs for the model are: (1) CME speed at $20 R_S$; (2) the drag parameter, $\gamma$; and (3) solar wind speed, $v_{sw}$. The parameter $\gamma$ is estimated by considering the relative density of the ejecta to the surrounding corona, as well as it's radial thickness. 
Its value, however, is not well constrained, and a heuristic rule given is that $\gamma = 10^{-8}$ km$^{-1}$ for bright CMEs and $\gamma = 2 \times 10^{-7}$ km$^{-1}$ for dim events (as inferred from coronagraph images), where a default value of $\gamma=0.2\times10^{-7}~\mathrm{km}^{-1}$ in combination with $v_{\mathrm{SW}}=450$ km s$^{-1}$ can be applied to a broad range of different CMEs. 
In addition, the DBM-ENLIL comparison indicated that although the DBM describes CME-ejecta propagation, it could also be applied as a proxy of the CME-shock propagation using lower values of the drag parameter, e.g. $\gamma=0.1\times10^{-7}~\mathrm{km}^{-1}$. \citet{vrsnak14a} compared real-time ENLIL forecasts until 2014 with hindsight runs of the DBM showing that both models performed similarly, with standard deviations of the predicted versus observed arrival times of about 14 hours.

To varying degrees, the remaining models in the CME scoreboard forecasts can be considered variants of these models, at least in terms of their attempt to identify some subset of solar observations that are used to drive a model, resulting in a prediction of the CME/shock arrival time at Earth. One model worth remarking on is the so-called ``average of all Methods''. This is an unweighted average of all the forecast submissions. As such, it represents a real-time ensemble prediction. 

All the teams that submitted forecasts to the CCMC are summarized in Table~\ref{table-models}. The number of forecasts produced by each group ranged from 1 to 114. Because different models were applied at different times, this resulted in a total of 139 possible forecasts, which explains why the ``average of all Methods'' entry has this number of forecasts. In the analysis that follows, we include the results from all of the models when we consider the statistics across all the models. However, for the purposes of investigating the properties of the forecasts in more detail, such as their variability from year to year, we restrict the analysis to the six most-frequently submitted models (``Average of all Methods'', ``WSA-ENLIL + Cone (GSFC SWRC)'', ``SIDC'', ``WSA-ENLIL + Cone (NOAA/SWPC)'',  ``WSA-ENLIL + Cone (Met Office)'', and `` Ensemble WSA-ENLIL + Cone (GSFC SWRC)''). We note that the results from the last model were not necessarily the ``official'' forecast. In some cases, results were inadvertently posted online from a preliminary, and not final forecast.  

\begin{table}[ht]
\caption{Summary of all models available on the CCMC's CME scoreboard website, including the number of forecasts performed by each model. Models are ordered by date of first submission. }
\label{table-models}
\centering
\begin{tabular}{lc}
\hline
Model Name & Number of Forecasts \\ 
\hline
WSA-ENLIL + Cone (GSFC SWRC)   &  114  \\ 
WSA-ENLIL + Cone (NOAA/SWPC)   &  78  \\ 
ips.gov.au   &  3  \\ 
 H3DMHD (HAFv.3+3DMHD)   &  1  \\ 
Anemomilos   &  18  \\ 
 Average of all Methods   &  139  \\ 
 ESA   &  3  \\ 
DBM   &  13  \\ 
BHV   &  4  \\ 
SIDC   &  101  \\ 
 STOA   &  8  \\ 
 HAFv2w   &  1  \\ 
Ensemble WSA-ENLIL + Cone (GSFC SWRC)   &  57  \\ 
 WSA-ENLIL + Cone   &  10  \\ 
 Expansion Speed Prediction Model   &  4  \\ 
COMESEP   &  7  \\ 
SARM   &  6  \\ 
 SAO Crowdsource   &  3  \\ 
WSA-ENLIL + Cone (Met Office)   &  70  \\ 
Rice-ENLIL Dst   &  1  \\ 
WSA-ENLIL + Cone (KSWC)   &  20  \\ 
 ElEvo   &  3  \\ 
SPM2   &  21  \\ 
WSA-ENLIL + Cone (BoM)   &  4  \\ 
SPM   &  14  \\ 
DBM + ESWF   &  3  \\ 
 EAM (Effective Acceleration Model)   &  8  \\ 
BGS   &  2  \\ 
Ooty IPS   &  2  \\ 
 NSSC SEPC   &  3  \\ 
 Other   &  2  \\ 
CAT-PUMA   &  1  \\ 
\hline
\end{tabular}
\end{table}

\subsection{Data}

Data were obtained from the CCMC's CME scoreboard website (\url{https://kauai.ccmc.gsfc.nasa.gov/CMEscoreboard}). These pages allow registered users to submit forecasts of CME-shock arrival times in real-time, compare with other submissions, and, once the event has arrived at 1 AU compare with the observed arrival time. The site takes as input: (1) predicted arrival time; (2) confidence in prediction; (3) Date and time of submission; and (4) Predicted geomagnetic storm parameters. The ``confidence in prediction'' parameter is a heuristic probability that the CME will actually be measured at 1 AU, with 0\% indicating that the ICME will definitely not be observed and 100\% indicating that it will certainly be observed. This may be of value in cases where the ICME trajectory with respect to Earth is anticipated to be glancing. From these, the website routines also add (1) the actual shock arrival time; and (2) the difference between the predicted and observed timing. Prediction of both CME-shock arrival time and geomagnetic parameter (Kp or $D_{st}$) is not required. 

The data are presented in HTML tables, with each year given on a separate web page. We wrote a java-based scraping routine to pull all the data from the current year page as well as the previous years (back through 2013), combine into a single structure, and write out a csv data file. This is provided in the supplemental information through a GitHub repository, and can be run at any point in the future to create an updated version of the dataset. 

\section{Results}

\begin{figure}[ht]
\centering
\includegraphics[width=30pc]{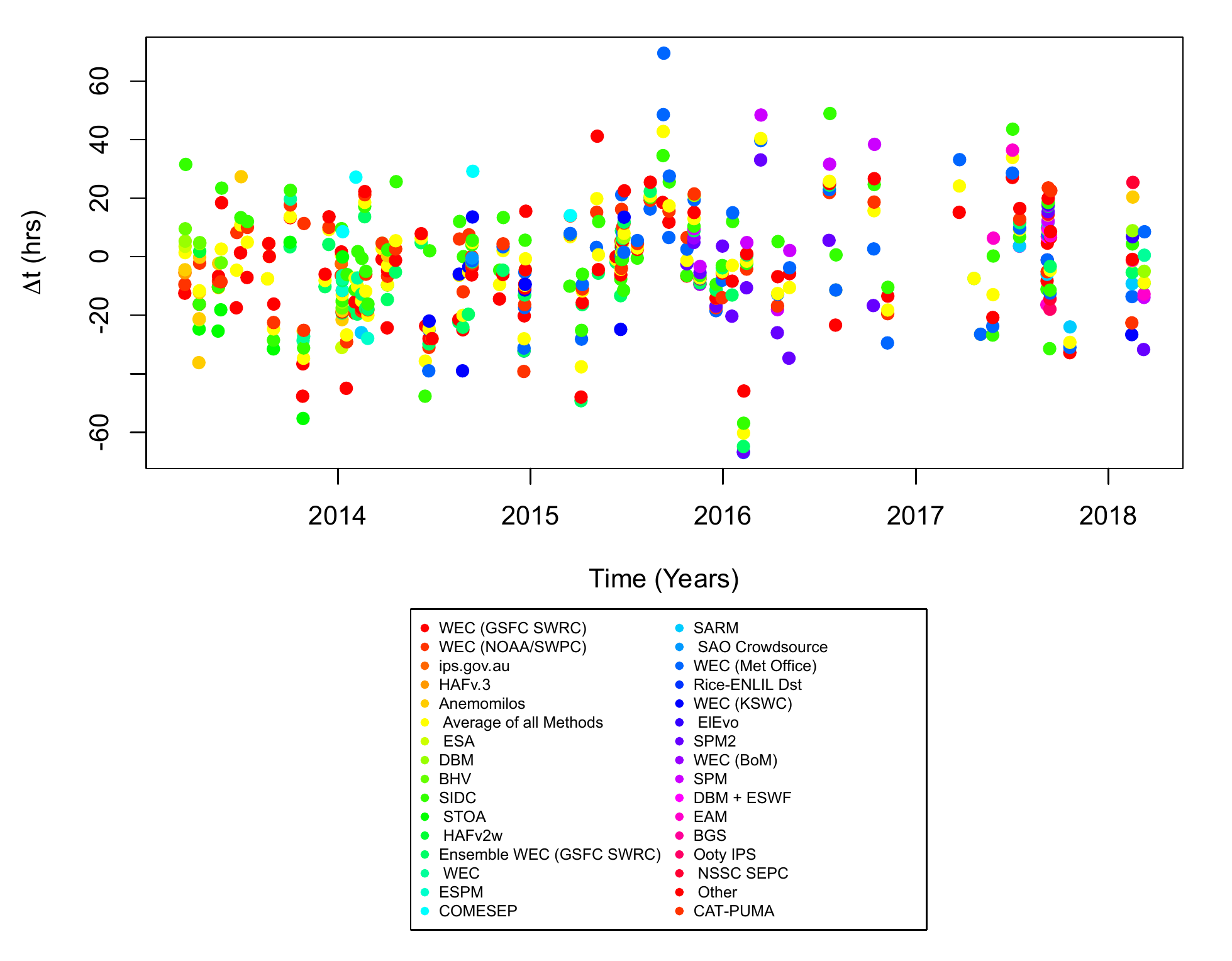}
\caption{Time series of all forecasts in the CCMC CME scoreboard database. Each circle has been color-coded according to the model that produced that prediction.}
\label{fig-all-forecasts-ts}
\end{figure}

As of May 11, 2018, there were 724 forecasts in the CME scoreboard forecast database. These were distributed amongst the models as shown in Table~\ref{table-models}. In Figure~\ref{fig-all-forecasts-ts}, we show $\Delta t$ as a function of time for all forecasts. The points have been color-coded according the model making that particular prediction. 

Figure~\ref{fig-all-forecasts-ts} summarizes all model predictions during the entire six-year period over which the CME scoreboard has been run.  Each model has been given a unique color to highlight any biases in prediction, as well as to indicate during which portion of the interval the model forecasts were being submitted. We note several points. First, at least qualitatively, there do not appear to be any obviously better models (as would be indicated by traces at, or near $\Delta t = 0$, although this is difficult to robustly assess from this display). Second, although there are no gross trends in the envelope (suggestive of a net improvement or worsening of the models), we note a tendency for forecasts around 2014 to be displaced below the zero-line, while no shift is apparent from mid-2016 onwards. Third, the number of forecasts appears to remain roughly constant during the interval. 

\begin{figure}[ht]
\centering
\includegraphics[width=30pc]{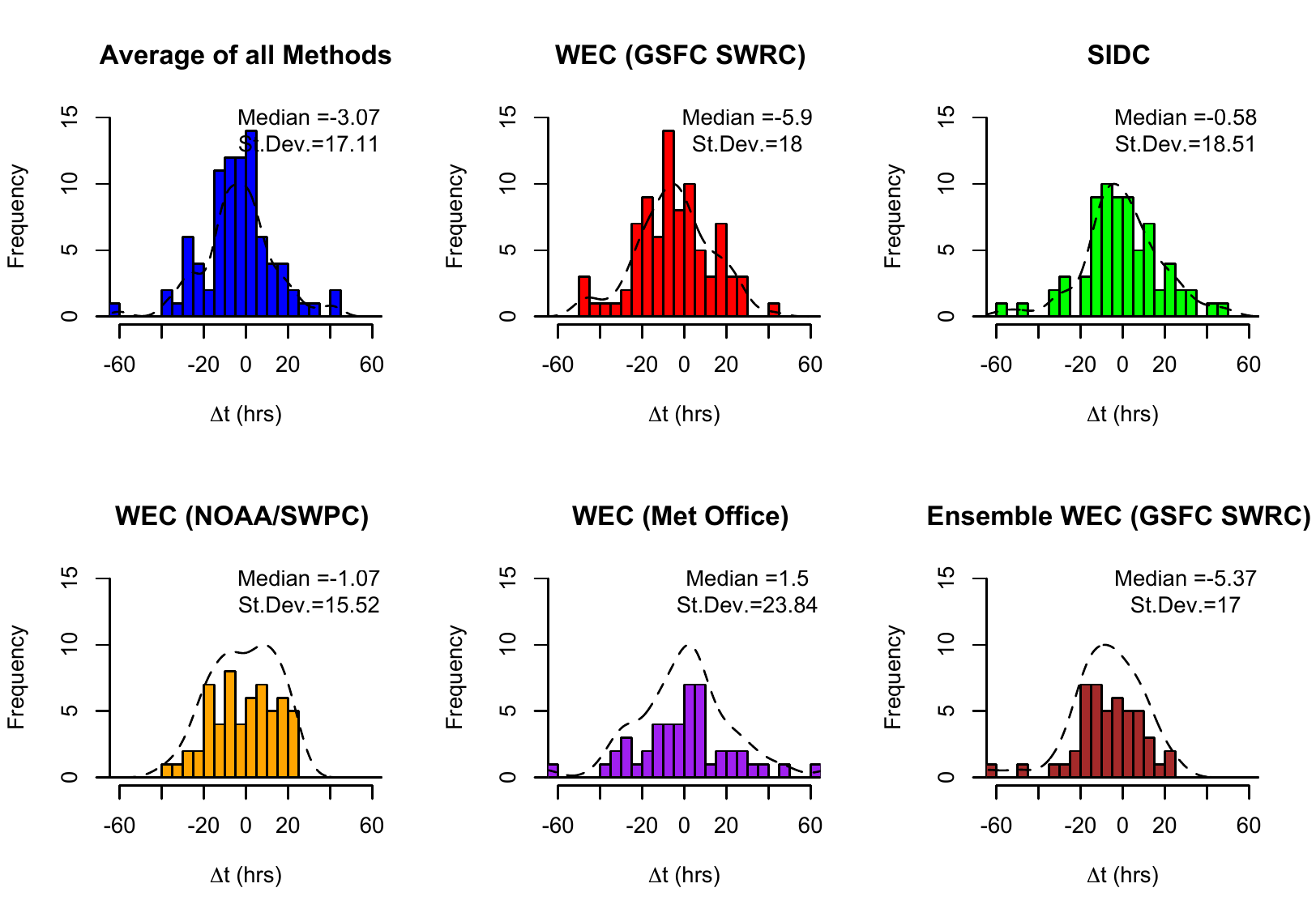}
\caption{Histogram of $\Delta t$ forecasts for six most-frequently submitted models. Values shown outside the x axis range summarize outliers, in this case, $< -60$ or $>60$ hours. WEC refers to the WSA-ENLIL + Cone model.}
\label{fig-top-six}
\end{figure}

In Figure~\ref{fig-top-six} we summarize the distribution of $\Delta t$ for the six most-frequently submitted models (i.e., where the ranking is based on the number of predictions, not necessarily the accuracy of the prediction). In each panel's title, the name of the model is given as well as the standard deviation in the prediction. The accuracy of the model is given by how far away from the true value the mean or median values lie. In these cases, the different models are reasonably clustered around zero. In contrast to accuracy, precision refers to the spread in the estimates (about the calculated average). Here, at least qualitatively, we see that the forecasts show similar degrees of spread. Table~\ref{table-top-six} provides more quantitative estimates of these quantities. Of note is that the SIDC forecasts are the only ones with mean and median errors that are less than one hour. The standard deviation, or spread in the distributions of five of the models is within 18.5 hours, with only the Met Office model having a slightly larger standard deviation. In what follows, we define model accuracy by $< \Delta t >$ and precision by standard deviation.  

\begin{figure}[ht]
\centering
\includegraphics[width=30pc]{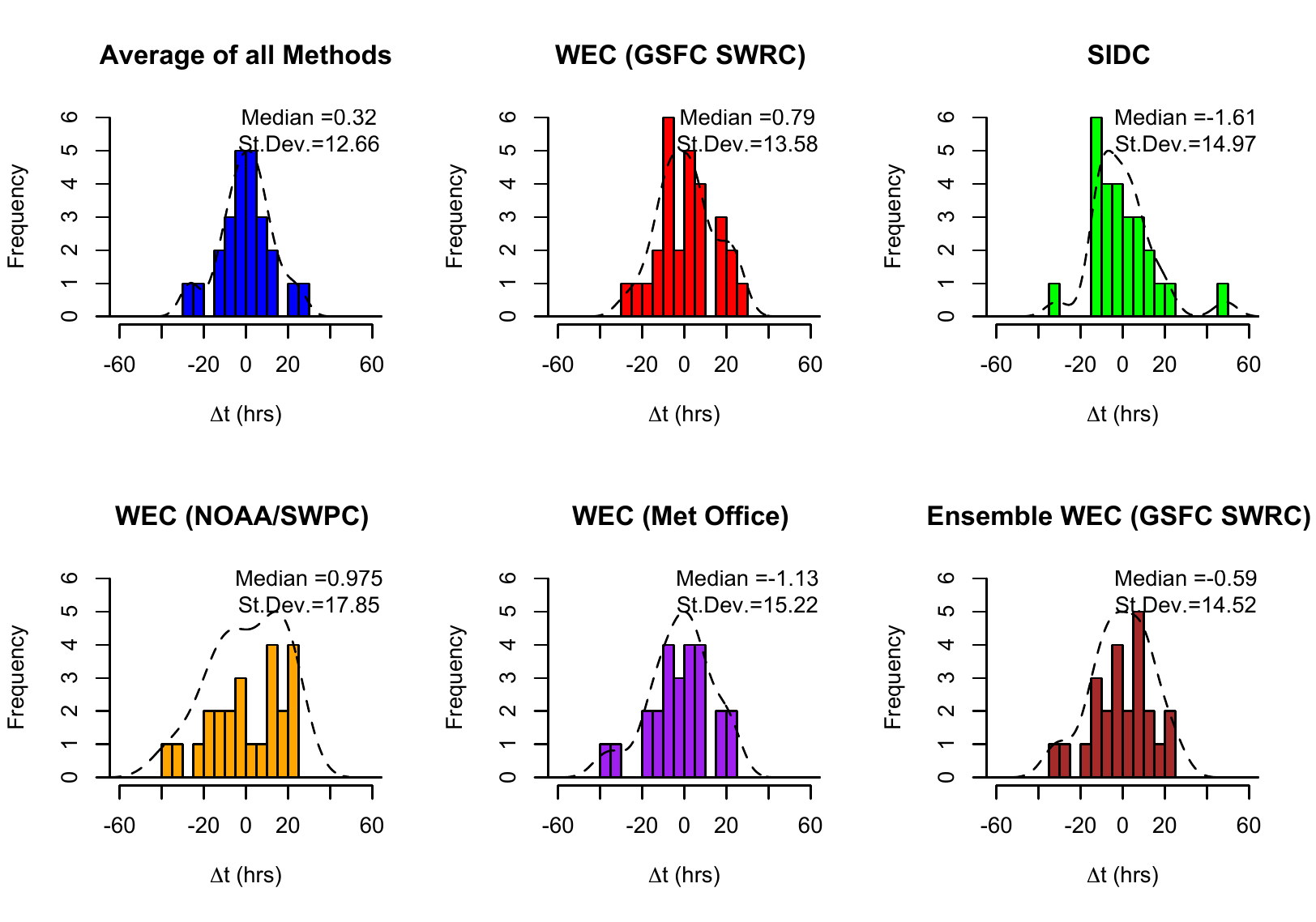}
\caption{Histogram of $\Delta t$ forecasts for the overlapping 28 CMEs of the six most-frequently submitted models.}
\label{fig-top-six-same-events}
\end{figure}

Table~\ref{table-stats-all-models} summarizes the main statistics for $\Delta t$ for all 32 models submitted to the CME scoreboard. We note the following main points. First, only one model (SIDC) had median and mean errors of less than one hour. Additionally, the team responsible submitted 101 forecasts, allowing us to conclude that this small offset are probably robust. Second, five models provided forecasts with mean and/or median offsets that were approximately 19 hours or more: STOA, ESA, COMESEP, Expansion Speed Prediction Model, and Rice-ENLIL Dst. Of these, four of them were associated with negative $\Delta t$'s, suggesting a strong bias to forecast a later arrival time than was actually observed. Third, in most cases the mean and median values were not significantly different, suggesting that the distribution of errors was relatively symmetric. In some cases, however, the 1st and 3rd quantiles were quite different; however, this was probably due to low-number statistics for a particular model, more than any intrinsic asymmetric distribution in the errors. Fourth, the MAE, although showing significant variation between the models is typically $\sim 10$ hours for models with a reasonably large number of submissions. For this subset of models, the ``Average of all Methods'' model slightly outperformed all others. 

\begin{sidewaystable}
\caption{Summary of the statistics for $\Delta t$ for all models. All times are in hours. $^{1}$ MAE refers to the Mean Absolute Error. $^{2}$ NA's occur when a model submits some subset of information for a forecast of an observed event, but which does not include an arrival time. $^{3}$ WEC refers to the WSA-ENLIL + Cone model. }
\label{table-stats-all-models}
\centering
\begin{tabular}{lccccccccccc}
\hline
Model  &  Min. & 1st Qu.  & Median  &  Mean & MAE & 3rd Qu.   & Max.  &  s.d. & L.T. & No. Forecasts  & NA's \\
\hline
WSA-ENLIL + Cone (GSFC SWRC)   &  -48  &  -15.8  &  -5.9  &  -4.89  &  14.5  &  5.65  &  41.2  &  18  &  55.94  &  114  &  30  \\ 
WSA-ENLIL + Cone (NOAA/SWPC)   &  -39.2  &  -11.9  &  -1.07  &  -1.27  &  13.1  &  11  &  23.5  &  15.5  &  52.15  &  78  &  20  \\ 
ips.gov.au   &  -5.47  &  -4.71  &  -3.95  &  -3.95  &  3.95  &  -3.19  &  -2.43  &  2.15  &  29.97  &  3  &  1  \\ 
 H3DMHD (HAFv.3+3DMHD)   &  -5.47  &  -5.47  &  -5.47  &  -5.47  &  5.47  &  -5.47  &  -5.47  &  NA  &  28.82  &  1  &  NA  \\ 
Anemomilos   &  -36.2  &  -19.2  &  -6.02  &  -6.35  &  14.4  &  -0.0675  &  27.3  &  17.8  &  40.36  &  18  &  6  \\ 
 Average of all Methods   &  -60.3  &  -11.7  &  -3.07  &  -3.67  &  12.9  &  4.82  &  42.8  &  17.1  &  NA  &  139  &  54  \\ 
 ESA   &  -31  &  -26.3  &  -21.5  &  -16.3  &  18.7  &  -9  &  3.53  &  17.9  &  37.57  &  3  &  NA  \\ 
DBM   &  -15.5  &  2.66  &  5.16  &  4.16  &  8.27  &  8.28  &  20.9  &  9.49  &  33.83  &  13  &  3  \\ 
BHV   &  -17.5  &  -5.94  &  1.36  &  -1.32  &  8.48  &  5.98  &  9.53  &  11.8  &  28.035  &  4  &  NA  \\ 
SIDC   &  -56.9  &  -10.2  &  -0.58  &  -0.129  &  13.6  &  11.2  &  48.9  &  18.5  &  54.4  &  101  &  30  \\ 
 STOA   &  -55.3  &  -28.5  &  -24.7  &  -21.5  &  22.9  &  -9.12  &  4.92  &  20.1  &  46.765  &  8  &  1  \\ 
 HAFv2w   &  1.8  &  1.8  &  1.8  &  1.8  &  1.8  &  1.8  &  1.8  &  NA  &  17.68  &  1  &  NA  \\ 
Ensemble WSA-ENLIL + Cone (GSFC SWRC)   &  -64.8  &  -17  &  -5.37  &  -7.12  &  13.8  &  4.48  &  24.4  &  17  &  40.07  &  57  &  10  \\ 
 WSA-ENLIL + Cone   &  -28.7  &  -12.4  &  -6.9  &  -7.08  &  12.2  &  0.372  &  19.7  &  15.6  &  46.3  &  10  &  2  \\ 
 Expansion Speed Prediction Model   &  -27.9  &  -23.8  &  -19.7  &  -19.7  &  19.7  &  -15.6  &  -11.5  &  11.6  &  32.5  &  4  &  2  \\ 
COMESEP   &  8.53  &  12.7  &  20.6  &  19.7  &  19.7  &  27.7  &  29.2  &  10  &  63.03  &  7  &  3  \\ 
SARM   &  -26  &  -20.3  &  -7.62  &  -8.27  &  13.5  &  1.23  &  11.9  &  14.9  &  49.005  &  6  &  NA  \\ 
 SAO Crowdsource   &  -2.5  &  -1.98  &  -1.46  &  -1.46  &  1.47  &  -0.948  &  -0.43  &  1.46  &  46.5  &  3  &  1  \\ 
WSA-ENLIL + Cone (Met Office)   &  -66.9  &  -12.6  &  1.5  &  0.237  &  17.3  &  9.14  &  69.5  &  23.8  &  46.375  &  70  &  23  \\ 
Rice-ENLIL Dst   &  -25  &  -25  &  -25  &  -25  &  25  &  -25  &  -25  &  NA  &  72.37  &  1  &  NA  \\ 
WSA-ENLIL + Cone (KSWC)   &  -39  &  -25.3  &  -10.4  &  -12.2  &  17  &  -2.15  &  13.6  &  17.1  &  50.925  &  20  &  8  \\ 
 ElEvo   &  4.85  &  5.84  &  6.83  &  9.89  &  9.89  &  12.4  &  18  &  7.09  &  9.97  &  3  &  NA  \\ 
SPM2   &  -66.7  &  -23.2  &  -10.6  &  -11.3  &  19.6  &  4.23  &  33  &  23.8  &  19.93  &  21  &  6  \\ 
WSA-ENLIL + Cone (BoM)   &  -9.4  &  -3.2  &  2.65  &  3.35  &  8.62  &  9.2  &  17.5  &  11.4  &  -5.025  &  4  &  NA  \\ 
SPM   &  -18.1  &  -1.96  &  7.18  &  11.2  &  18.2  &  26.8  &  48.4  &  22  &  19.6  &  14  &  4  \\ 
DBM + ESWF   &  6.87  &  7.53  &  8.18  &  8.18  &  8.19  &  8.84  &  9.5  &  1.86  &  17.5  &  3  &  1  \\ 
 EAM (Effective Acceleration Model)   &  -16.4  &  -4.12  &  4.95  &  4.35  &  11.9  &  8.29  &  36.4  &  16.3  &  35.185  &  8  &  NA  \\ 
BGS   &  6.87  &  6.87  &  6.87  &  6.87  &  6.87  &  6.87  &  6.87  &  0  &  11.085  &  2  &  NA  \\ 
Ooty IPS   &  -17.9  &  -11.6  &  -5.22  &  -5.22  &  12.7  &  1.14  &  7.5  &  18  &  17.475  &  2  &  NA  \\ 
 NSSC SEPC   &  -14.4  &  0.035  &  14.5  &  8.48  &  18.1  &  19.9  &  25.4  &  20.6  &  34.53  &  3  &  NA  \\ 
 Other   &  -7.5  &  -7.5  &  -7.5  &  -7.5  &  7.5  &  -7.5  &  -7.5  &  NA  &  22.19  &  2  &  1  \\ 
CAT-PUMA   &  -5.33  &  -5.33  &  -5.33  &  -5.33  &  5.33  &  -5.33  &  -5.33  &  NA  &  43.47  &  1  &  NA  \\ 
\hline
\end{tabular}
\end{sidewaystable}

These data, however, do not account for any possible biases in the selection of which events the forecasters choose to forecast. Some approaches, for example, may be suited to predictions of certain types of events, or teams may have only participated in the scoreboard during particular intervals. Thus, restricting our analysis to only those events that were forecasted by all models would be more instructive in assessing the intrinsic skills of the modelers and not any inherent predictability of the subset of ICMEs considered by each team. Since many models only provided a handful of forecasts, we limited this comparison to the six models that submitted the most forecasts. From these, we identified 28 events that were forecasted by all six. 
Figure~\ref{fig-top-six-same-events} summarizes the distribution of  $\Delta t$ for these events. Again, the dashed line provides an estimate of the density function, providing some measure for how Gaussian-like these distributions may be. 
We note the following points. First, the values are generally distributed more normally than when all events are considered, with the ``Average of all Methods'' being most Gaussian. Second, the events are generally more narrowly constrained, suggesting that this subset of events were more amenable to prediction. Third, the median values are all very small, with the ``Average of all Methods'' performing best. The standard deviations were all similar, again with the ``Average of all Methods'' yielding the lowest value. 

\begin{figure}[ht]
\centering
\includegraphics[width=30pc]{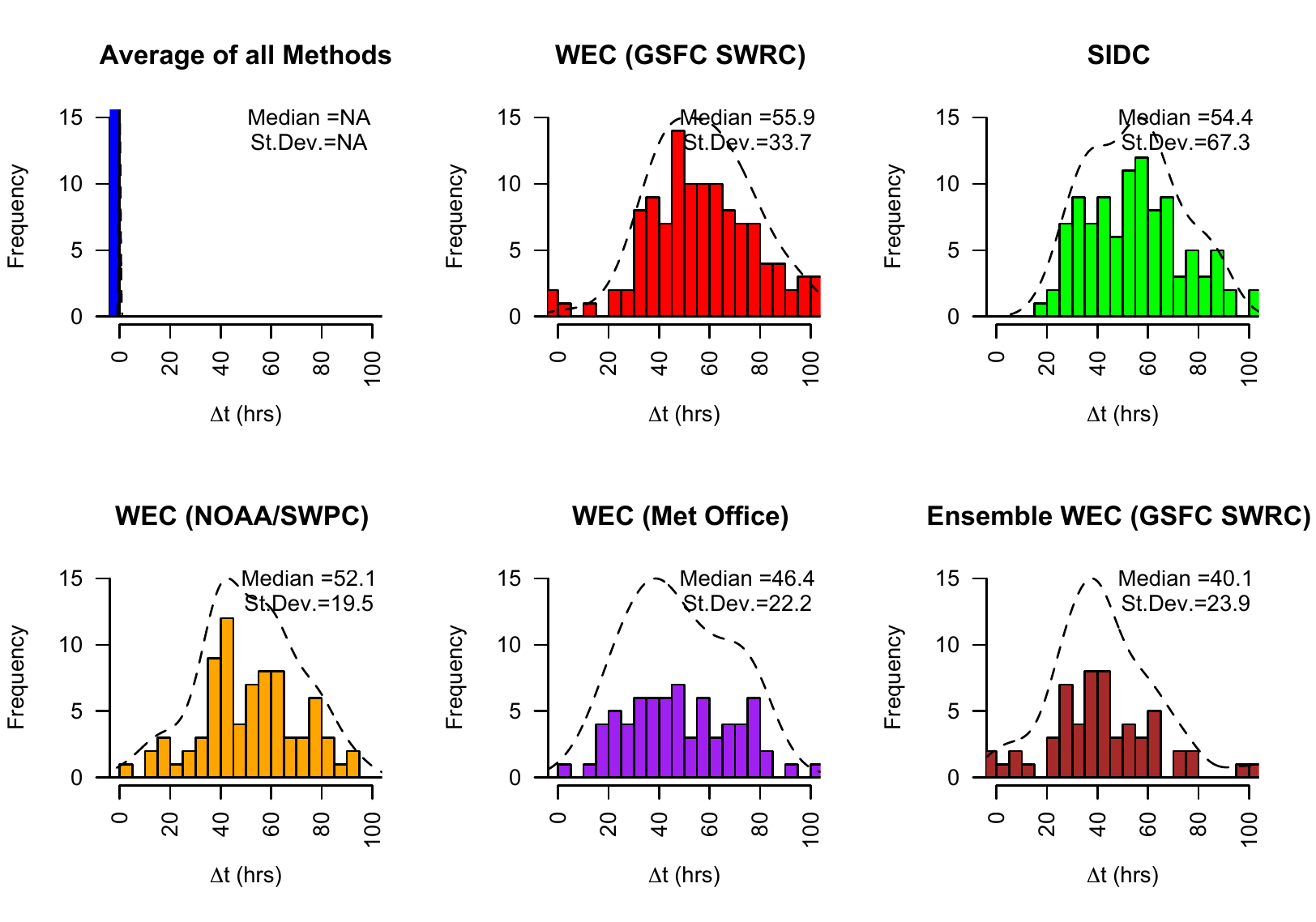}
\caption{Histogram of lead time for all forecasted CMEs for six most-frequently submitted models.}
\label{fig-top-six-lead}
\end{figure}

The CME scoreboard table also provides the lead time for each forecast, i.e., the difference between the observed arrival time of the CME shock and the time the forecast was submitted to the CCMC. These are summarized in Figure~\ref{fig-top-six-lead} for all ICMEs that were forecasted by the previously-defined six models. In this case, no lead time was provided for the ``average of all methods'' since it represents the average of multiple forecasts: We could have computed the average of all lead times for the submitted forecasts, but, it could be argued that the lead time should be the smallest of all values, since the forecast could not be computed until that last prediction was submitted. For the remaining models, the median and spread in values are approximately the same, with SIDC having the largest standard deviation (due, at least in part, to one large outlier). The distribution of the Met Office's model's lead times are flatter, with a lower median value than other WEC approaches. Finally, we note that two submissions from the WEC (GSFC SWRC) team had negative lead times, corresponding to forecasts that were made (or at least submitted) after the ICME arrived at 1 AU. 

\begin{figure}[ht]
\centering
\includegraphics[width=30pc]{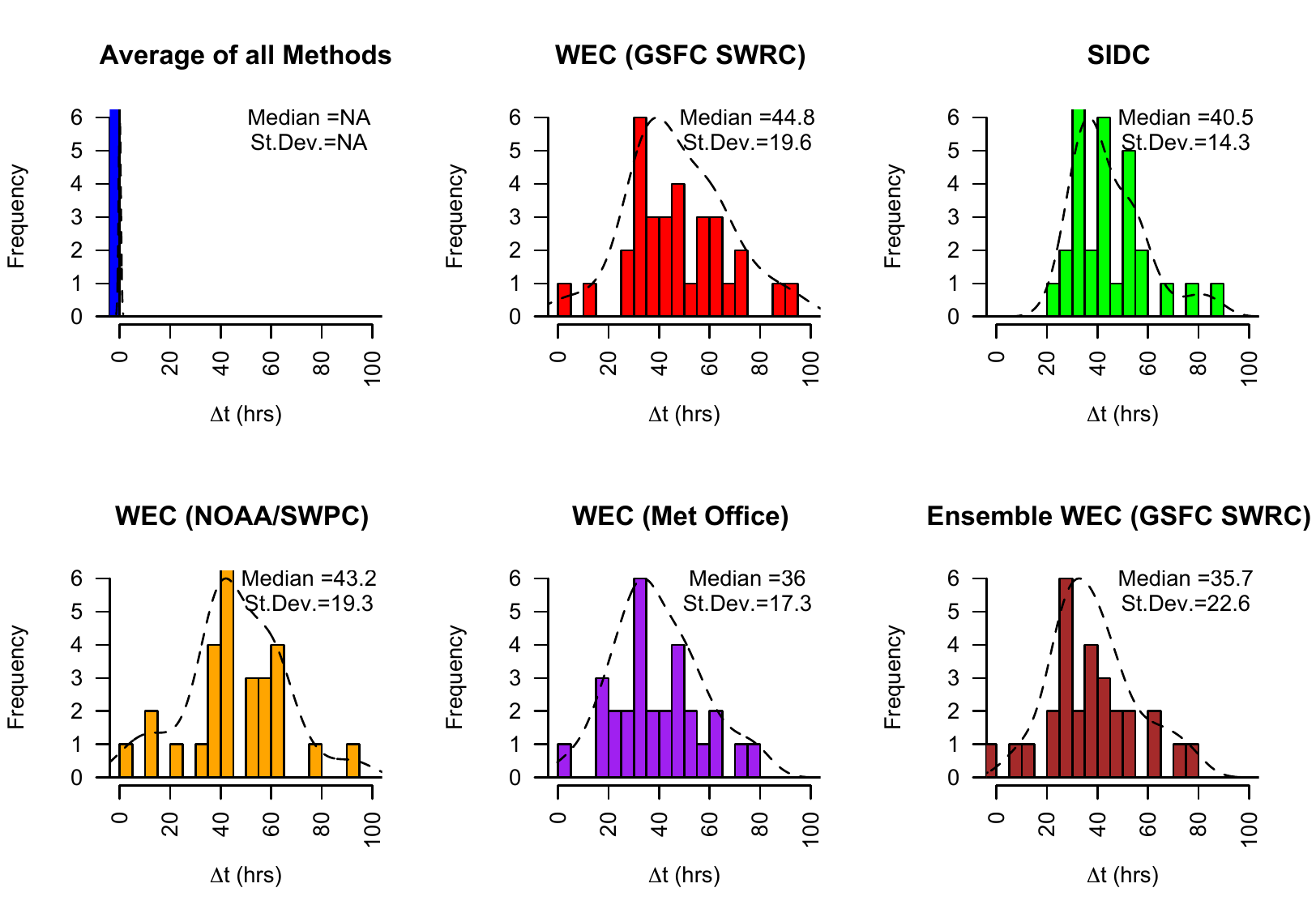}
\caption{Histogram of lead times for overlapping 28 CMEs for six most-frequently submitted models.}
\label{fig-top-six-lead-same}
\end{figure}

Again, restricting ourselves to the 28 events that were sampled by the six most frequently submitted models, the picture changes (Figure~\ref{fig-top-six-lead-same}). Now the lead times all drop moderately to approximately 40 hours. The spread also decreases and the distributions look more log-normal-like. For these events, predictions were consistently made with lead times greater than 20 hours.  

\begin{table}[ht]
\caption{Summary of the statistics for $\Delta t$ for the six most-frequently submitted models. All times are in hours. $^{1}$ MAE refers to the Mean Absolute Error. $^{2}$ WEC refers to the WSA-ENLIL + Cone model.}
\label{table-top-six}
\centering
\begin{tabular}{lcccccccc}
\hline
Model &   Min. & 1st Qu. &  Median &   Mean & MAE$^{1}$ & 3rd Qu.  &  Max. & s.d. \\
 \\ \hline
 Average of all Methods   &  -60.3  &  -11.7  &  -3.07  &  -3.67  &  12.9  &  4.82  &  42.8  &  17.1  \\ 
WEC$^{2}$ (GSFC SWRC)   &  -48  &  -15.8  &  -5.9  &  -4.89  &  14.5  &  5.65  &  41.2  &  18  \\ 
SIDC   &  -56.9  &  -10.2  &  -0.58  &  -0.129  &  13.6  &  11.2  &  48.9  &  18.5  \\ 
WEC (NOAA/SWPC)   &  -39.2  &  -11.9  &  -1.07  &  -1.27  &  13.1  &  11  &  23.5  &  15.5  \\ 
WEC (Met Office)   &  -66.9  &  -12.6  &  1.5  &  0.237  &  17.3  &  9.14  &  69.5  &  23.8  \\ 
Ensemble WEC (GSFC SWRC)   &  -64.8  &  -17  &  -5.37  &  -7.12  &  13.8  &  4.48  &  24.4  &  17  \\ 
\hline
\end{tabular}
\end{table}

\begin{figure}[ht]
\centering
\includegraphics[width=30pc]{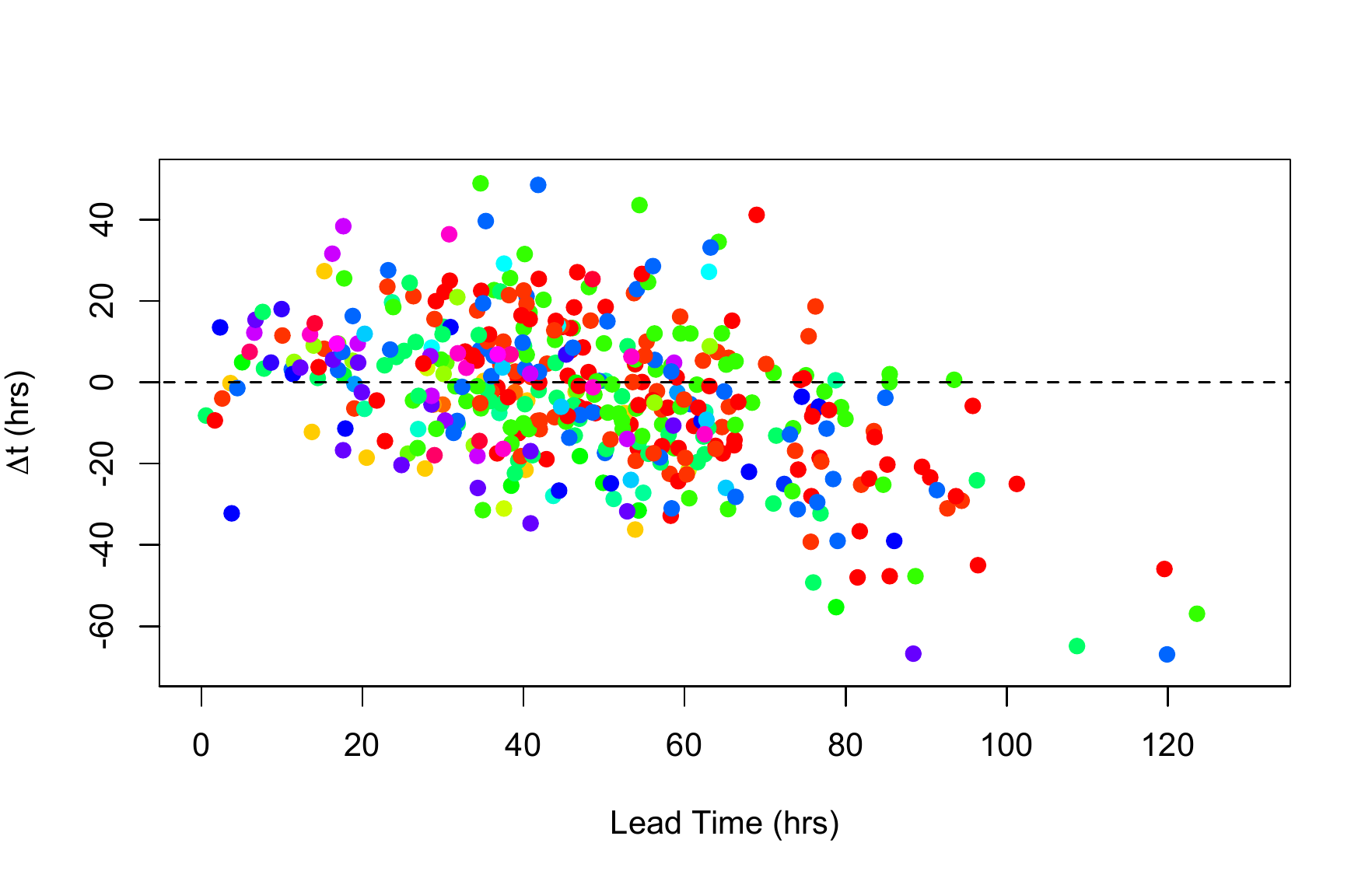}
\caption{Comparison of lead time versus arrival time error for all CME forecasts. The color-coding is the same as that in Figure~\ref{fig-all-forecasts-ts}}
\label{fig-lead-time-vs-dt-all}
\end{figure}

Figure~\ref{fig-lead-time-vs-dt-all} makes a comparison between $\Delta t$ and Lead time for all forecasts made during the six-year period. The data have been color-coded as in Figure~\ref{fig-all-forecasts-ts}. There appears to be a general trend that as the lead time of the forecast increases, the errors in arrival time are negatively biased. That is, that the models predict an earlier arrival time than was observed. A similar, but slightly different interpretation is that for lead times less than 70 hours, there is no obvious bias in arrival time; however, beyond 70 hours, the forecasts are strongly negatively biased. While this display provides gross trends across all submissions, it is difficult to make any inferences on the performance of specific models. Although we have not included external data in our analysis of the CME scoreboard, it could be reasonably inferred that the longer-lead-time events are associated with slower CMEs (since they take longer to propagate to Earth), and, thus, the bias in the longer-lead-time events represents errors introduced in trying to forecast slow CMEs. 

\begin{figure}[ht]
\centering
\includegraphics[width=30pc]{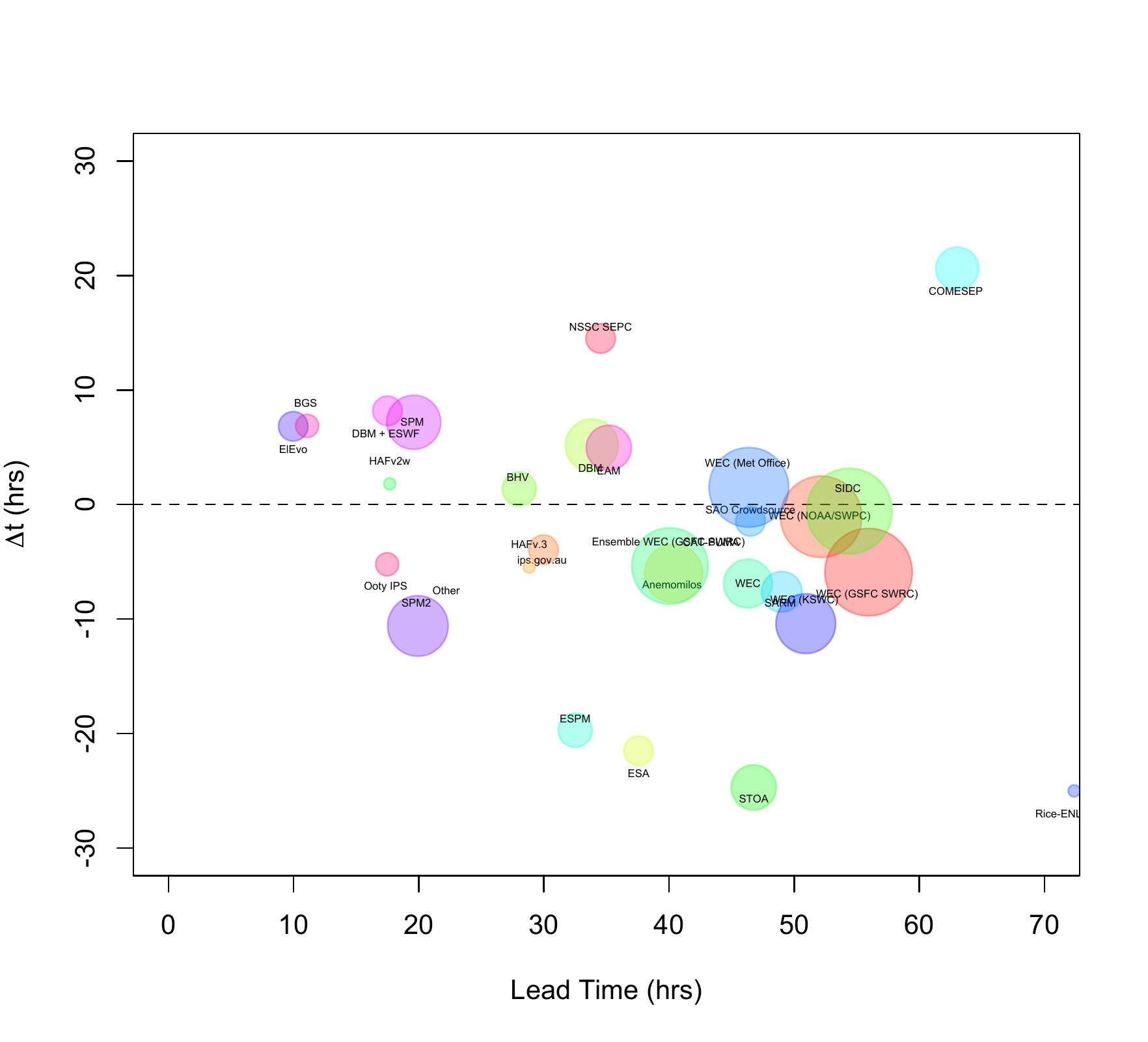}
\caption{Comparison of median lead time versus median arrival time error for forecast model as a function of time. The size of the bubbles is proportional to the logarithm of the number of forecasts made by that model, with a lower threshold size set so that models with only a few forecasts still remain visible.}
\label{lead-time-vs-delta-t-bubble}
\end{figure}

Figure~\ref{lead-time-vs-delta-t-bubble} addresses this. Again, $\Delta t$ is plotted against lead time; however, in this case all forecasts by a specific team have been averaged and plotted as one point, with the radius of the bubble being proportional (logarithmically) to the number of forecasts made by that team. We note several points. First, the trends observed in Figure~\ref{fig-lead-time-vs-dt-all} are not as apparent. The WEC-based models are generally associated with longer-lead-time forecasts than many of the other models. Second, the teams submitting the most forecasts are all clustered around $\Delta t \sim 0$, with most of them lying slightly negative. Third, the SIDC, NOAA/SWPC, and Met Office models are notably the best performers in having the least bias, with SIDC slightly outperforming the other two given (1) a smaller $\Delta t$ and (2) longer lead time. 

\begin{figure}[ht]
\centering
\includegraphics[width=30pc]{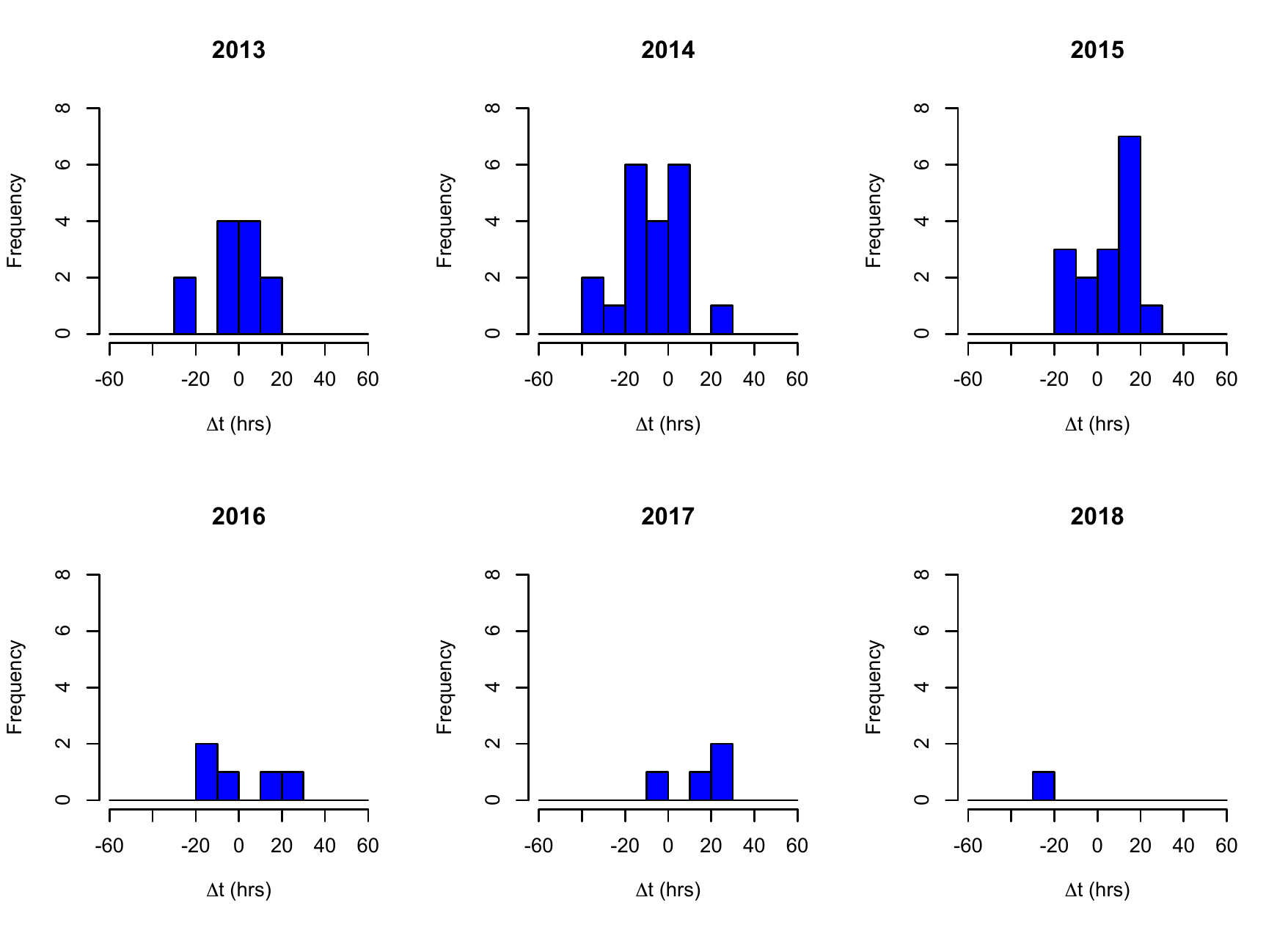}
\caption{Histogram of $\Delta t$ from NOAA/SWPC forecasts, based on the WSA-ENLIL+Cone Model as a function of year.}
\label{fig-swpc-by-year}
\end{figure}

Turning our attention now to possible temporal trends, in Figure~\ref{fig-swpc-by-year} we show the distribution of $\Delta t$ from the NOAA/SWPC model as a function of time (in years). Similar results were found for SIDC (results not shown). The statistics inferred from these values are summarized in Table~\ref{table-swpc}. We note the following points. First, there is a significant variation in the total number of forecasts each year, the first and last years perhaps representing truncation constraints. The largest number of forecasts were made in 2014 and 2015. Second, there do not appear to be any obvious asymmetries: Some years show a slightly larger positive tail (e.g., 2016), while other years show a negative tail (e.g., 2015). Third, the MAE shows no obvious statistical trend moving from 2013 to 2017. Heuristically, at least, the MAE appears to increase as time progresses. 

\begin{figure}[ht]
\centering
\includegraphics[width=20pc]{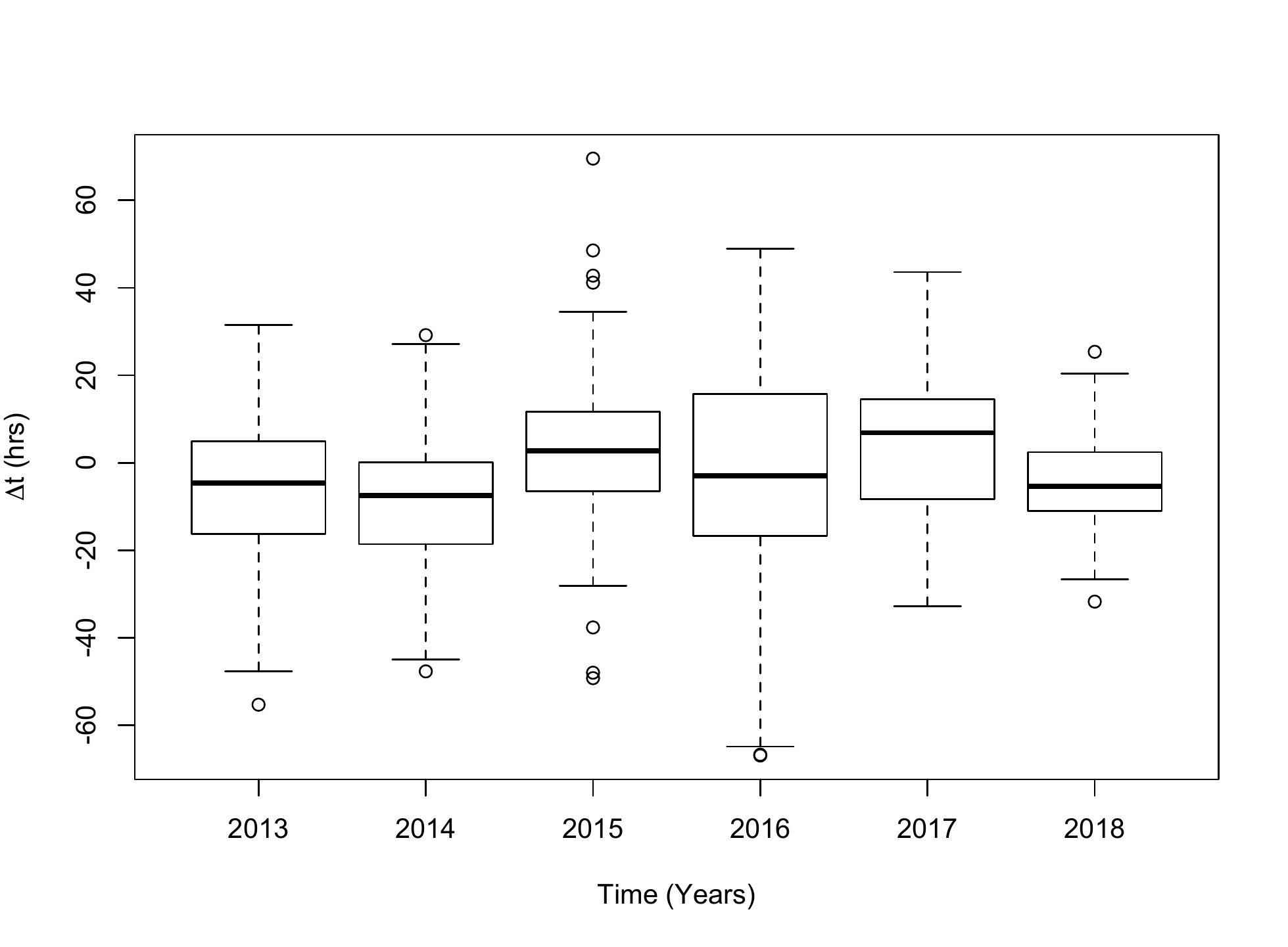}
\caption{Variation in $\Delta t$ as a function of year for all model predictions. The line in the centre of the box gives the median of the data, while the tops and the bottoms of the box give the lower and upper quartiles. The ends of the vertical lines give the minimum and maximum values of the data (provided that there are no outliers), while any circles give the values of outliers (more then 1.5 times above/below the upper/lower quartiles).}
\label{fig-whisker-all}
\end{figure}

\begin{table}[ht]
\caption{Summary of the statistics for $\Delta t$ for the NOAA/SWPC model by year. All times are in hours. }
\label{table-swpc}
\centering
\begin{tabular}{cccccccccc}
\hline
Year  &  Min. & 1st Qu.  & Median  &  Mean & MAE & 3rd Qu.   & Max.  &  s.d. & No. Forecasts \\
\hline
2013  &  -25.2  &  -8.8  &  -0.435  &  -1.47  &  11.2  &  10  &  17.7  &  13.7  &  17   \\ 
2014  &  -39.2  &  -18.3  &  -6.55  &  -8.53  &  13.2  &  3.07  &  21.2  &  14.9  &  26   \\ 
2015  &  -17.4  &  -5.34  &  8.31  &  5.22  &  12.2  &  15.3  &  21.4  &  12.6  &  23   \\ 
2016  &  -19.5  &  -16.8  &  -4.25  &  0.004  &  16.2  &  18.6  &  21.9  &  19.4  &  6   \\ 
2017  &  -5.13  &  8.3  &  17.7  &  13.4  &  16  &  22.8  &  23.5  &  13.3  &  5   \\ 
2018  &  -22.6  &  -22.6  &  -22.6  &  -22.6  &  22.6  &  -22.6  &  -22.6  &  NA  &  1   \\ \hline
\end{tabular}
\end{table}

Given the limitations of small-number statistics for analyzing a single model, we can ask whether, as a whole, the forecasts from all models have improved over the last six years. 
Figure~\ref{fig-whisker-all} shows the distribution of $\Delta t$ for all forecastss as a function of year in the form of a whisker plot. 
We note the following: (1) The number of forecasts rose for three years, then decreased by a factor of two for the last two years; (2) The median values bracket zero, suggesting that there is no obvious systematic bias in the forecasts; and (3) There is no obvious decrease (or increase) in the size of the boxes or maximum values during the almost six-year period. With respect to the first point, this likely represents two competing effects: As the CME scoreboard gained in popularity with modelers, the number of submitted forecasts increased; however, as the Sun has moved ever closer to solar minimum, the number of opportunities to make forecasts decreased. This can be inferred from the number of forecasts made by each model (See Table~\ref{table-stats-all-models}), and can also be inferred by the vertical clustering of the points, which appears to increase, or remain roughly constant moving from 2013 to 2017.

\section{Discussion}

In this study, we have investigated the accuracy of space weather models in forecasting the arrival time of CMEs and/or their associated shocks. Taken as a whole, the models can, on average, predict arrival times to within $\pm 10$ hours, however, the precision around this average is large: $\pm 20$ hours. These results compare well with those of \citet{wold17a}, who found a MAE in arrival time of $10.4 \pm 0.9$ hours for 273 CMEs predicted (and observed) to arrive at Earth, STEREO A or STEREO B. 
On average, the best performers can predict the arrival time to within one hour (mean error), with a MAE of 13 hours, and a standard deviation in these predictions of $<15$ hours. Since the mean error was generally negative for the six most-frequently submitted models, this suggests a forecasted arrival time later than observed, and, hence, a systematic bias in the forecasts. The mean error is notably smaller than the early prediction error of $\sim 4.0$ hours reported by \citet{wold17a}. It is also worth noting that all of these predictions are made in real-time, and were initiated by remote solar observations. Thus, the lead time in the forecasts is substantial: Typically the time it takes for the CME to travel from $20-30 R_S$ to 1 AU. For a fast CME, traveling at an average speed of 1,000 km/s, this would translate into a lead time of 38 hours. 

The ``Average of all Methods'' forecasts generally performed as well as, or outperformed the other models. As an unweighted average of all forecast times, it represents a simple super-ensemble approach. Its performance is somewhat surprising given the large errors from many of the submitted forecasts. However, by relying on a basic tenant of ensemble modeling - that random errors from different models tend to cancel in the averaging procedure - it is able to achieve excellent forecasts. Its primary limitation lies in the fact that it cannot be calculated until all submissions have been received. Thus the effective lead time for the ``Average of all Methods'' is governed by the date/time of the final submission. 

Forecasting the arrival time of CMEs and/or their shocks represents a pragmatic decision. On one hand, it is an easily defined and intuitive metric, and is relatively straightforward to estimate. On the other hand, it is not necessarily the crucial piece of information about the event we want to know. For example, if $B_z$ is forecasted to become strong and southward, and remain so for a prolonged period, the precise timing of this is of secondary importance \citep[e.g.][]{savani15a,kubicka16a,kay17a,riley17b}. Similarly, if you can predict that $B_z$ will remain zero, or only show positive excursions, then the timing of this matters little. By extension, forecasting $v_{SW}$, Kp, $D_{st}$, and AE, even with poor knowledge of the exact timing may be of considerably more value to users of space weather products.  In spite of this, predicting time of arrival also provides a path for incrementally improving space weather models. For example, for one of the variants of the WSA-ENLIL+Cone model to improve substantially, which would likely be in improving the precision of the forecast, it likely requires some improvement to the CME model used (currently a simple plasma ejection), the specification of the ambient solar wind (currently based on an empirical relationship), or more sophisticated treatment of multiple eruptions (which appear to play an important role in the largest CME-driven storms). Additionally, as shown by \citet{mays15a,mays15b}, improving the accuracy in the arrival time of an ICME improves the accuracy of the speed of the CME, which, in turn, improves the prediction of Kp (if solar wind speed is used to predict Kp). 

The arrival time of CME-driven shocks depends on a number of factors that could be used to further refine our analysis. For example, the speed (initial, average, or final) of the CME likely affects the overall accuracy of the forecast: Fast and massive CMEs arrive more quickly, and are not decelerated by the background solar wind as much as slower events. Together these suggest that the uncertainties in arrival time would be lower. 

It is notable that the errors in arrival time amongst the various models were quite similar, both in terms of the accuracy and precision of the forecasts. The models represent a broad range of techniques and ideas for inferring the time taken to travel from the solar corona to 1 AU. Alternatively, they represent different methodologies for deriving the speed profile of the CME as a function of heliocentric distance. Each model is driven by a different, but usually overlapping set of input data: Flare time, type II radio bursts, or white light images, magnetograms, to name but a few. With each set of input data comes a different set of assumptions and approximations. In white light, for example, halo CMEs provide a measure of the initial speed of the ejecta, but, unless multiple views are available, this is a projected speed that, additionally, may reflect more of the expansion of the CME than its propagation toward the observer \citep{owens04a}. The profiles of $v(r)$ either assumed or computed by the model can also have a substantial effect on the arrival time. In principle, the WSA-ENLIL+Cone models offer the most accurate way to estimate this since the global model includes an ambient solar wind as background, and propagates the ejecta through it. However, while ambient solar wind models can, and often do reproduce the bulk features of the solar wind \citep{riley01a}, for any specific solar rotation, there can be substantial differences \citep{riley12e,owens07a, jian15a}. These differences can result in significant differences in the arrival time for CMEs. Moreover, in these physics-based models, no account is made of the magnetic structure within the ejecta. The magnetic forces neglected because of this approximation can have an important effect on the evolution of the ejecta during its passage to 1 AU. 

In this study, we focused on the differences between the predicted and observed arrival times, and thus considered only events for which a prediction was made and this resulted in an observed shock at 1 AU, i.e., ``hits''. We did not consider: (1) ``false alarms'', where a model predicts a shock to arrive at a particular time, but no shock is subsequently observed; or (2) ``misses'', where a model made no prediction (e.g., because the model (or modeler) predicted that the shock would miss the Earth) but a transient shock was observed.  These are, obviously important scenarios to consider from a forecasting perspective, and skill scores could be defined to capture model capabilities for them \citep[e.g.][]{wold17a}. The CCMC team are currently developing a mechanism to track misses (that is, ICMEs that are measured in situ at 1 AU that no model predicted). 

On average, the best model predicted CME-driven shock arrival times to within an hour (mean error). However, as a metric, this may not be the most appropriate quantity to consider, since positive and negative errors tend to cancel one another. It is only when the direction of the difference is removed, such as with the MAE ($\sim 15$ hours), or when we consider the spread in these predictions (s.d. $\sim 15$ hours) that we can appreciate the need to improve the forecasts. Generally, as scientific understanding is transitioned to operational forecasts, it is typically the statistical, empirical, or ``heuristic'' models that outperform the mechanistic, or physics-based models; at least in the beginning. Here, however, the best model is also the most complex, incorporating the most sophisticated numerical techniques and relevant physical processes. Thus, the best opportunities to improve upon these forecasts probably lie in mitigating any uncertainties and errors within the WSA-ENLIL+cone model, several of which have been noted above. Here, we suggest several other ideas that may improve the forecasts. First, the specification of the ambient solar wind (the ``WSA'' part of the model) relies on a questionable prescription for the solar wind speed \citep{riley15a}. If the ``expansion factor'' model is replaced by the ``distance from the coronal hole boundary'' prescription, this may result in smaller errors. 

Second, the magnetograms used to drive the background solar wind flow are plagued with uncertainty \citep{riley14a}. Synoptic maps, that is, maps built up from 27-day observations from Earth, or near-Earth space, can differ significantly from one solar observatory to another. These can translate into substantial differences in the predicted solar wind speed at 1 AU \citep{riley10d}, which, in turn, can significantly modulate the speed of an ICME as it propagates from the Sun to Earth, and hence, affect its arrival time. 

Third, the uncertainties in the specification of the ejecta must be improved (the ``Cone'' part of the model). In \citet{mays15a}, a considerable part of the final arrival-time error was caused by minor variations in the initial direction and speed of the cone. This is particularly true for the most energetic CMEs, for which a large fraction of the corona participates in the eruption. Additionally, Heliospheric Imagers (HI), which allow the tracking of CMEs along the entire Sun-Earth line \citep[e.g.][]{davis09a}. \citet{mostl17a}, using HI images from STEREO A and a self-similar expansion (SSEF) method, showed that for a set of 76 Earth impacting CMEs, the mean error in accuracy was $3 \pm 16$ hr. For a smaller dataset, \citet{wood17a} estimated the uncertainties in the arrival time of 28 well observed ICMEs (identified in the Wind in-situ measurements and remote solar observations), concluding that the s.d in arrival time was  11.7  hours, reducing to 6.3 hours by the time the ICME had reached 0.3 AU. Although these numbers cannot be compared directly with the results of the CCMC CME scoreboard, which were estimated in real-time,  the narrowing of the error by almost a factor of two suggests that improvements in the interpretation and/or fitting of the near-Sun observations can have a significant effect on the precision of the forecasts. Additionally, it provides support for an operational spacecraft situated at L5, which would be able to continuously track a CME from the Sun to the vicinity of Earth. By 0.3 AU, the speed of the ejecta may not change dramatically. Thus, this is likely the optimum location to maximize both the lead time of the prediction and its accuracy \citep{colaninno13a}. Ultimately, HI data contain information about the location of the CME, which should improve the accuracy of the arrival time of the CME, at the expense of a shorter lead time (as compared to the models described here, which use initial conditions derived from coronagraph data only (for future prospects with HI, see also \citet{harrison17a}). Ideally, a combination of HI data with improved numerical simulations in real time would seem a reasonable avenue for making progression.

A related issue concerns the shape of the CME front and/or shock ahead of it. \citet{kubicka16a} used the DBM to constrain the kinematics of a CME that was observed by both coronagraphs and at Venus (0.72 AU). Since Venus was only $\sim 6^{\circ}$ away from the Sun-Earth line, it had been anticipated that the estimated arrival time would be substantially more accurately determined. The error, however, was 6 hours. They suggested that ICME-ICME interactions could have played a role in modulating the speed of the CME, but it is also likely that strong, local curvatures in the CME/shock front at least contributed to the error in arrival time. Thus, any transverse local inhomogeneities in the CME front close to the Sun -- which provide the initial conditions for the modeled ejecta -- would be magnified as it evolves during its journey to 1 AU. 

It is worth remarking that forecaster ``error'' or ``bias'' is an under-appreciated but potentially significant source of error in the forecasts. 
In at least several teams (e.g., SIDC, Met Office, and NOAA/SWPC), the forecaster uses model predictions as a guidance for a much broader forecast, which takes into account many other contributions, including, but not limited to: the number and quality of available coronal images; the confidence in the CME fit, the presence of preceding events that might affect the medium into which the ICME is propagating, as well as the confidence the forecaster has in the WSA map prescribing ambient conditions. 
Additionally, the people making the forecasts have likely changed over time. NOAA/SWPC's predictions, for example, were originally made by two researchers (D. Biesecker and G. Milward), however, later, they were replaced by operational forecast personnel. Interestingly, there is no obvious change (for better or worse) coincident with the replacement of personnel.

In this study, we did not attempt to convolve forecast lead-time with the accuracy of the prediction. However, the two quantities should be combined to produce a more meaningful metric. Longer lead times, up to the time when the eruption is first observed should, generally, be viewed as more valuable than forecasts made closer to when the CME reaches Earth, since the user community has more time to implement any necessary mitigation strategies. However, it is not clear how to accomplish this, particularly, as its appeal will be domain dependent. Ideally, a perfect forecast (i.e., one where $\Delta t = 0$) made when the CME (or flare) is first observed, should receive a perfect skill score (although, in principle, we might build in the possibility that models can predict CME arrival times prior to eruption). A forecast that predicts a CME arrival time accurately at the time of arrival, or a model that predicts an event at flare time that does not intercept Earth probably has little to no merit. But how do we compare forecasts made at flare onset with $\pm10$ hour accuracy with forecasts made as the event crosses, say, 0.5 AU with $\pm5$ hour accuracy?   It is worth noting that for most events, and for most forecasts, forecasters submitted only one entry. However, in a few cases, double submissions were made by the same team, with different lead times. Unfortunately, there weren't a sufficient number of cases to assess whether forecasts improved as lead time decreased. 

This study has been a first step in assessing forecasts for the arrival times of CMEs at Earth; however, it is by no means a definitive assessment. We suggest several potentially fruitful avenues for continuing this work, which would likely involve the collection of data not currently available on the current CME scoreboard table. 
First, for purposes of formally ranking the models against each other (which was not a primarily goal for the current investigation), statistical tests, such as the Mann-Whitney or Wilcoxon signed rank test, could be used to compare each pair of models to show that for all the CMEs they both forecasted, one model predicted them better than the other. Second, as noted above, ``false alarms'' or missed forecasts could be included in the analysis through the construction of contingency tables \citep[e.g.][]{bloomfield12a}. Third, more sophisticated metrics for assessing arrival time could be constructed, such as one that combines lead-time with arrival-time. Fourth, a valuable investigation could centre on separating out automatic forecasts from those that rely on human-added elements \citep[e.g.][]{murray17a}. 

The limitations of the CCMC CME scoreboard also suggest some areas for improvement. For example, all teams should publish their official forecasts in a consistent, agreed-upon, and usable manner. Following the lead from terrestrial weather forecasting, official centres, in particular, should agree on what to publish, where to publish it, and how the assessment of the forecasts should be made (i.e., metrics) \citep{henley17a}. Additionally, publishing these publicly will provide the research community with crucial benchmarks with which to test other, novel approaches. 

In closing, we note that one or more of the teams submitting infrequent predictions to the CCMC may be capable of providing considerably more accurate forecasts. A limiting factor for many is the small number of events to which the model has been applied. As more forecasts are made, by these or other novel models, we may see new best performers. By using the code included with this report, we hope that the modelers will rigorously test and compare their results (for both past and future events) in an effort to improve accuracy. Additionally, we hope that the process of refinement would be simpler as multiple variants of each model (with selected parameters and/or inputs modified) can be quickly tested and compared with previous iterations.

\acknowledgments
We thank the forecasting teams for submitting their predictions for the arrival times of CME-driven shocks at various points over the last six years. In particular, we would like to thank the UK Met Office research, verification and forecast teams for detailed and constructive comments on the manuscript. PR and JL would like to thank NASA's Living with a Star Program for supporting this research (grant no. NNX15AF39G). TA and CM thank the Austrian Science Fund (FWF): [P26174-N27].  C. V. was funded by the Research Foundation - Flanders, FWO PhD fellowship no. 11ZZ216N.

\listofchanges

\end{document}